\documentclass[final]{aipproc}

\layoutstyle{6x9}

\begin{document}

\title{Understanding Nucleons in the Nuclear Medium}


\author{D. W. Higinbotham}{
  address={Jefferson Lab, Newport News, VA 23606, USA}
}

\author{V. Sulkosky}{
  address={Massachusetts Institute of Technology, Cambridge, MA 02139, USA}
  }

\begin{abstract}
Recent cross section (e,e'pN) short-range correlation experiments have clearly shown 
the strong dominance of tensor correlations for (e,e'p) missing momenta greater 
than the Fermi momentum; while recent $^2$H(e,e'p)n and $^4$He(e,e'p)t 
asymmetry experiments at low missing momentum have shown small changes from 
the free nucleon form factor.  By doing asymmetry 
experiments as a function of missing momentum, these results can be linked together 
and observed as a change of sign in the measured asymmetry. 
This idea will be presented within the context of the recently 
completed Jefferson Lab Hall~A quasi-elastic, 
polarized $^3$He(e,e'N) experiments (N=0,p,n,d) where the asymmetries of several 
reaction channels were measured with three, orthogonal target-spin directions. 
Together, these various experiments will help us to better understand 
nucleons in the nuclear medium.
\end{abstract}

\maketitle


\section{Introduction}

\begin{figure}[htb]
  \label{higinbotham-md}
  \includegraphics[height=.5\textheight]{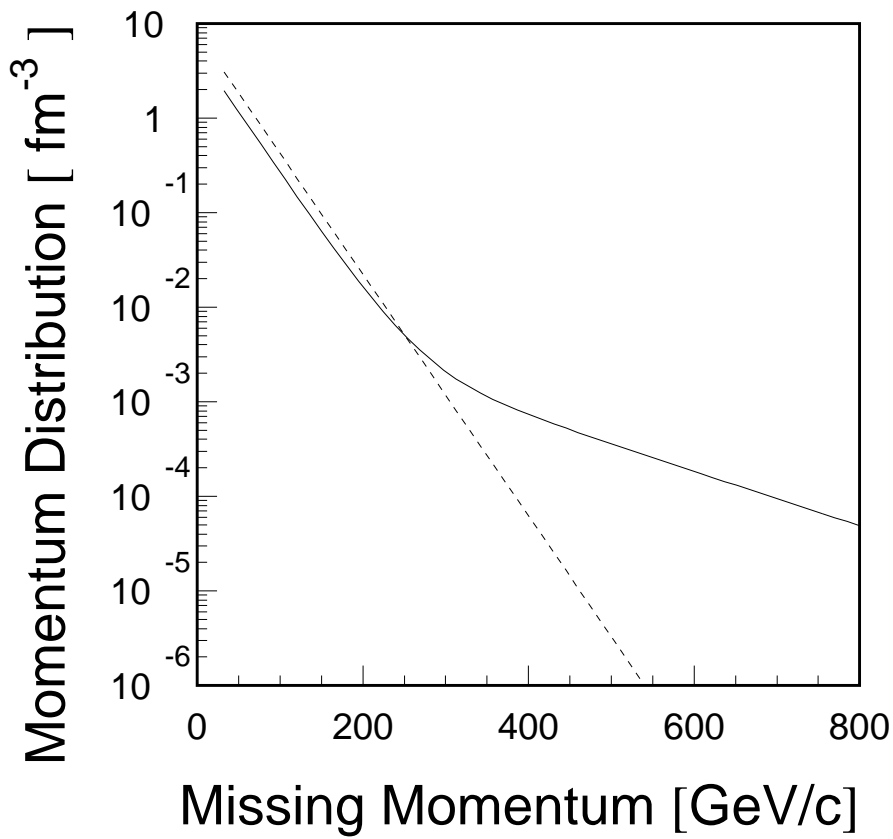}
  \caption{The solid line is an approximate momentum distribution of the nucleon in $^3$He based on the data
  from Benmokhtar {\it et al.}~\cite{Benmokhtar:2004fs}.  The inflection in the solid line above the 
  Fermi momentum (approximately 250~MeV/c) is due to tensor and short-range correlations between the nucleons in 
the nucleons.   The dashed line shows an independent-particle model distribution,
  which is larger at low momentum but continues to fall-off  exponentially at high momentum.}
\end{figure}

Early independent particle model calculations, while nicely predicting the inelastic energy levels in
nuclei, dramatically over predicted the occupation density of these states as measured in 
(e,e'p) experiments~\cite{Lapikas:1003zz,Kelly:1996hd}.
A spectroscopic factor of 60-70\% was needed to bring the model and the data into
agreement.  The simplest explanation for the discrepancy was that something
that doesn't effect the general shell-structure was taking strength 
away from the predicted states.  This is generally agreed to be caused by
the correlations between the nucleons in the nucleus and is illustrated in 
Fig.~\ref{higinbotham-md}.  As is shown, since the overall strength must be a constant, 
the effect of these correlations is to decrease the strength at low missing momentum and increase
the strength at high missing momentum relative to a independent-particle model.

Experimentally, the signature of a nucleon-nucleon correlations should be seen as knocked-out, 
back-to-back nucleons with relative momenta much greater then the Fermi momentum of around 250~MeV/c.
Unfortunately, many reaction mechanisms such as Meson Exchange Currents (MEC) 
and/or Final State Interactions (FSI) have this same final-state and thus complicate trying to
probe for initial-state correlations.  While Jefferson Lab's higher beam energy and high Q$^2$ 
was thought to be a solution to this
problem, the initial Jefferson Lab knock-out reaction data from two- and few-body systems 
at high missing-momentum were also sensitive to these reactions 
mechanisms~\cite{Benmokhtar:2004fs,Ulmer:2002jn,Rvachev:2004yr,Egiyan:2007qj}, though it was also
clear that corrections were a needed ingredient to explain the observed cross sections.

\section{Recent Progress}

\subsection{Cross Section Measurements}

Recent pair knock-out measurements have observed a very strong dominance of proton-neutron pairs
over proton-proton pairs.  This was first seen at Brookhaven where a proton beam was used as the 
probe~\cite{Piasetzky:2006ai} and then at Jefferson Lab
where an electron beam was used as the probe~\cite{Shneor:2007tu,Subedi:2008zz}.
Unlike most observables, the dominance of back-to-back high-momentum proton-neutron pairs can 
be explained neither by MEC nor FSI but instead is attributed by several theory groups to initial-state tensor 
correlations~\cite{Sargsian:2005ru,Schiavilla:2006xx,Alvioli:2007zz,Wiringa:2008dn}; thus, this observation 
is finally a clean signature of an initial-state correlations.

The effect of these same correlations has also become clear in inclusive scattering.
This is done by making ratios of inclusive (e,e') cross section as a function $x_B$
with $x_B$ = Q$^2$/2m$\omega$ where Q$^2$ is the square of the four-momentum transferred to the 
system, m is the mass of a nucleon, and $\omega$ is the energy transfer.
For Q$^2$ greater than 1~[GeV/c]$^2$, one can observe clear scaling in the $x_B > 1$ 
cross section ratio; a phenomenon known as y-scaling.  This was originally reported in
SLAC data~\cite{Frankfurt:1993sp} and later with higher precision CLAS 
inclusive data~\cite{Egiyan:2003vg}.  By taking the ratios to $^3$He instead of deuterium, the CLAS experiments
were able to not only show two-nucleon correlations from a scaling in the $1 < x_B < 2$ region~\cite{Egiyan:2003vg}, 
but also possible three-nucleon correlations by showing a second scaling in the $2 < x_B < 3$ region~\cite{Egiyan:2007qj}.

\subsection{Asymmetry Measurements}

As one considers these correlations in terms of an exactly calculable two-body system, the strong tensor correlations
are associated with the d-state of the deuteron's wave-function.  As pointed out by 
Friar~\cite{Friar:1979zz}, wave functions are not observables; but if one clearly defines the
basis, it is possible to calculate a d-state probability for a given momentum distribution.
In this spirit, it has been clearly shown within a model, such as Arenh\"{o}vel  {\it et al.}'s~\cite{Arenhoevel:1992xu},
one can predict the polarized beam-target asymmetry for proton knock-out from vector polarized deuterons
as a function of missing momentum.  Such models and experiments have shown that at low missing momentum the asymmetry 
is primarily due to the s-state of the nucleon-nucleon potential used in the model, whereas the high missing momentum 
asymmetry is primarily due to the d-state  of the nucleon-nucleon potential used in the model~\cite{Passchier:2001uc}.
Polarization transfer experiments done at low missing momentum on the $^4$He(e,e'p)$^3$H reaction have also 
not seen a large effect at low missing momentum~\cite{Strauch:2002wu}.

\section{Three-Body System}

At Mainz, the $^3$He(e,e'pn)p cross section has recently been measured and 
again shows the dominance of the proton-nucleon reaction (tensor-correlations)
in the momentum range above the Fermi momentum~\cite{Middleton:2009rf}.  Interestingly, the Faddeev calculations
using AV-18 or Bonn-B nucleon-nucleon potentials over-estimate the magnitude of the $^3$He(e,e'pn)p 
cross section at low missing momentum.    While from this experiment alone it is hard to say why there is a 
discrepancy, it is clear that there is still more physics going on then in these very up-to-data calculations. 

Complimenting this new cross section result, an extensive series of quasi-elastic polarized $^3$He
measurements have been made at Jefferson Lab for a Q$^2$ of 0.2, 0.5 and 1.0~[GeV/c]$^2$ 
with the polarized target's spin oriented in three orthogonal directions~\cite{e05015,e05102,e08005}.  
These experiments used the two Hall~A High Resolution Spectrometers (HRS)~\cite{Alcorn:2004sb}, 
the BigBite magnet~\cite{NUIMA.A406.182,NUIMA.A412.254} with a customized hadron detector package 
for identification of protons and deuterons, and the 3~m high, 0.4~m deep, and 1~m wide Hall~A Neutron 
Detector (HAND) that was used for the $^{12}$C(e,e'pn) experiment~\cite{Subedi:2008zz}.  With 
this combination of detectors, it was possible to measure the (e,e'), (e,e'p), (e,e'n), and (e,e'd) 
data simultaneously.   This experiment was completed during the summer of the 2009, and analysis 
is now underway.  By measuring so many reaction channels over a large range in Q$^2$, it is hoped 
that we will be able to dramatically improve our understanding of the three-body system.


\clearpage
\begin{theacknowledgments}
We thank Misak Sargasian (Florida International University), Mark Strikman (Penn State University) and 
Eli Piasetzky (Tel Aviv University) for many interesting and enlightening discussions 
on the subject of short-range correlations.  We also thank the hard work of the four Ph.D. students
working on the analysis of the Jefferson Lab Hall~A polarized $^{3}$He data: 
Jin Ge (University of Virginia), 
Ellie Long (Kent State University), 
Miha Mihovilovic (University of Ljubljana). and
Yawei Zhang (Rutgers University).
\end{theacknowledgments}

\bibliographystyle{aipproc}   
\bibliography{hadron2009-src}

\end{document}